# Developing a Quantitative Resiliency Approach

Vincent P. Paglioni[1*], Graeme Troxell[2], Aaron Brown[3], Steve Conrad[2], Mazdak Arabi[4]

[1]*Risk, Reliability, & Resiliency Characterization (R3C) Lab, Department of Systems Engineering, Colorado State University. Fort Collins, CO, USA.*
[2]*Blue Green Decisions Lab, Department of Systems Engineering, Colorado State University. Fort Collins, CO, USA.*
[3]*Department of Systems Engineering, Colorado State University. Fort Collins, CO, USA.*
[4]*Department of Civil and Environmental Engineering, Colorado State University. Fort Collins, CO, USA.*

**Abstract**. Resiliency has garnered attention in the management of critical infrastructure as a metric of system performance, but there are significant roadblocks to its implementation in a realistic decision-making framework. Contrasted to risk and reliability, which have robust quantification approaches and undergird many regulatory approaches to system safety (e.g., "risk-informed decision-making"), resiliency is a diffuse, qualitatively-understood characteristic, often treated differently or distinctly. However, in the emerging context of highly-complex, highly-interdependent critical systems, the idea of reliability (as the probability of non-failure) may not be an appropriate metric of system health. As a result, focus is shifting towards resiliency-centered approaches that value the *response* to failure as much as the avoidance of failure. Supporting this approach requires a robustly-defined, quantitative understanding of resiliency. In this paper, we explore the foundations of reliability and resiliency engineering, and propose an approach to *resiliency-informed decision-making* bolstered by a quantitative understanding of resiliency.

## 1. Introduction

Reliability and resiliency are both fundamental attributes of physical system performance and safety. These attributes inform our understanding of how systems behave over time, particularly under stress or failure conditions, yet their analyses are frequently siloed. Furthermore, reliability and resiliency capture distinct aspects of system performance, meaning a siloed treatment fails to capture holistic system behaviors. Reliability, as the probability of component (or system, or mission, etc.) non-failure under stated conditions for a given period of time (Modarres & Groth, 2023), has been well-defined for even complex systems, and risk (as a proxy for the complement of reliability) has been incorporated into risk-informed practices to safeguard system operations (U.S. Nuclear Regulatory Commission, 2009). However, neither risk nor reliability, with the focus on avoidance and occurrence of failures, respectively, consider system *response* to failure in qualification or quantification – thus neglecting a significant portion of system safety behavior. Resiliency, despite varying definitions, encompasses both the preparation for and, critically, *response to* system failures.

As systems are becoming more complex, interweaving traditionally distinct domains of hardware, software, communications, and human operations, understanding and ensuring system reliability is an increasingly challenging endeavor (Aalund & Paglioni, 2025a). Furthermore, the interconnection of critical systems is creating highly-complex fault propagation networks, as evidenced by, e.g., the CrowdStrike outages in 2024 (O'Flaherty, 2024), further complicating the assessment of system reliability. The emerging context of extreme system and inter-system complexity calls into question the applicability of reliability as a suitable metric for system performance – that is, begging the question of whether "non-failure" is a sufficient basis for engineering decisions. Instead, as we propose herein, the basis for system-level decision-making should be shifted to value *both* the avoidance of and response to failure. That is, *resiliency*, not simply reliability, should be the basis of engineering decision-making.

Despite reflecting different facets of system robustness (and often being treated distinctly), reliability and resiliency are deeply interconnected (V. P. Paglioni et al., 2025). As will be shown, resiliency has two main facets:

- **Proactive resiliency** identifies system components and behaviors that contribute to planning for and anticipating perturbations
- **Reactive resiliency** encompasses system behaviors that contribute to absorbing, recovering, and/or adapting during and/or after the occurrence of perturbations

The two-facet framing of resiliency inherently connects it to reliability, which is similar to proactive resiliency. Reliability, as the probability of non-failure, captures the system behaviors and components that contribute to avoiding perturbations. The avoidance of perturbations is one aspect of anticipation, and so reliability is necessary, albeit not sufficient, for describing proactive system resiliency.

A resilient system is designed to survive and recover from perturbations. As discussed, this includes reliability as a piece of proactive resiliency. Figure 1 displays the relationships between reliability and resiliency as a Venn diagram.

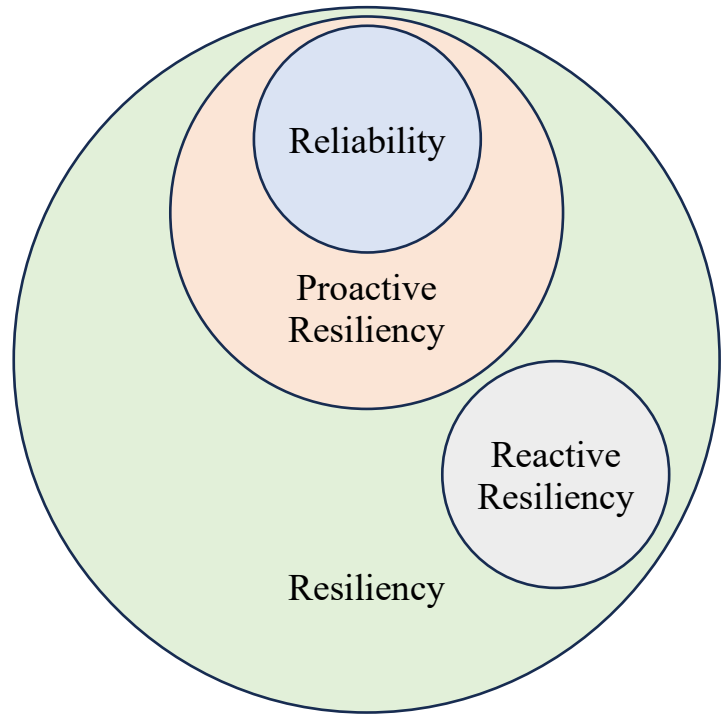

Figure 1. Connection between system reliability and resiliency.

Despite the parallels in aspects of reliability and resiliency, however, there are also conflicts induced in the pursuit of reliability or resiliency as a "paramount" system characteristic:

- A system designed for maximum reliability may be optimized for normal operations and thus highly efficient, but unprepared for, and thus brittle in the face of, unmodeled/unexpected events. Highly reliable systems may lack the capacity to adapt, reroute, or self-heal, thereby possessing low resiliency.
- Conversely, a highly resilient system may tolerate frequent disturbances or component-level failures gracefully, reconfiguring itself or recovering with minimal disruption. However, this flexibility can reduce the system's reliability if it entails greater complexity or more interacting parts.

This interplay reveals a critical insight: reliability must fail in order for resiliency, particularly reactive resiliency, to become visible as a system characteristic. Thus, reliability and resiliency operate on different timelines. Reliability is often the focus during normal operations (i.e., before a perturbation), while resiliency dominates the response to a disruption in normal operations. Clearly, both of these behaviors must be considered to fully characterize system behavior over its lifecycle.

Despite their overlap, reliability and resiliency are often evaluated separately. Partially, this is out of extrinsic necessity – the *pursuit* of system reliability emphasizes failure probabilities, mean time between failures (MTBF), and probabilistic risk assessments. The pursuit of avoiding failure necessarily focuses on pre-failure conditions and how to minimize failure, at the cost of neglecting post-failure system behaviors; how it copes, adapts, or recovers. On the other hand, resiliency emphasizes post-failure dynamics without necessarily accounting for *how* the failure occurred. The response time, recovery effectiveness, degradation tolerance, etc. can be somewhat viewed in isolation from the occurrence of the initiating failure. Some resiliency approaches are qualitative, and may not quantify failure likelihood at all. These disparate approaches lead to gaps in understanding as a system may be assumed to be inherently resilient or unrealistically binary (working or failed). That is, in either focus, system behavior is idealized and realistic scenarios may be missed.

This paper seeks to emphasize the value in both reliability and resiliency, and the insights facilitated when approaches are combined, rather than pursued in isolation. Before implementing a unified approach, however, resiliency needs to be robustly and cohesively defined for engineered systems, and implemented within a quantitative causal modeling architecture. A holistic approach, as will be shown, co-designs systems for reliability – as necessary to proactive resiliency – and reactive resiliency, to avoid the pitfalls associated with the single-minded pursuit of either characteristic. This paper lays the groundwork for a unified approach to reliability and resiliency and the implementation of *resiliency*-informed decision-making for complex systems, especially critical infrastructure.

Critical infrastructure, encompassing water distribution, power generation and distribution, communications, and transportation systems (among many others) is perhaps the quintessence of the increasingly complex and interconnected system paradigm. Failures in these systems arise from novel and varied threat vectors, propagate across increasingly tightly-coupled connections, and induce new, dynamic intra- and inter-system behaviors. Furthermore, these systems form the core of the modern global economy and quality of life. It is imperative that these systems be resilient, and so the context for initially defining and quantifying resiliency is centered around critical infrastructure.

The remainder of this study is structured as follows. Section 2 reviews risk, reliability, and resiliency as separate constructs in engineering. Section 3 discusses the tradeoffs between reliability and resiliency approaches, as well as the synergies connecting these two metrics that can be leveraged to improve the practice of both. Section 4 presents a first approach to bridging reliability and resiliency through the lens of repairable system modeling techniques (reliability) extended to model resiliency actions. Section 5 discusses the broader implications of this approach to resiliency, and Section 6 concludes with the future work required to fully develop resiliency as an engineering discipline.

## 2. Background: Risk, Reliability, and Resiliency

Risk, reliability, and resiliency are critical system characteristics that, despite capturing different behaviors, are deeply interconnected. This section provides the necessary background to contextualize the remainder of this work. In the following subsections, each characteristic is defined and explained in isolation. Section 3 starts the discussion of bridging reliability and resiliency.

### 2.1. Risk

Risk can be defined at different levels. Broadly, a *risk* is the possibility of unwanted consequences. That is, risk in this context is both inherently uncertain and inherently negative. Engineering risk is rooted in the "risk triplet" (Kaplan & Garrick, 1981), which defines the three components of risk:

1. *Scenario:* What accident sequences are possible in the system?
2. *Consequence:* What is the result of each accident sequence?
3. *Probability:* How likely is each accident sequence to occur?

Thus, system risk can be identified at the level of individual threats (single accident sequence) or amalgamated across threat vectors (all accident sequences) to identify the "total system risk." This approach treats the scenario as an implicit element of risk and essentially the "structure" around which risk is quantified. The risk $F$ of a single scenario $s_i$ is obtained by Equation (1) as the product of scenario consequence $C(s_i)$ and probability $\Pr(s_i)$.

$$F(s_i) = C(s_i) \times \Pr(s_i) \tag{1}$$

The risk of the system is then the sum of risks across all considered scenarios, as shown in Equation (2).

$$F(sys) = \sum_{i=1}^{n} F(s_i) = \sum_{i=1}^{n} C(s_i) \times \Pr(s_i) \tag{2}$$

Risk captures the probability and consequences of adverse events occurring at *some* point in a system's life, but the characterization of risk in Equations (1) and (2) does not explicitly address the temporal aspect of risk. Even when graphed using typical approaches like Event Trees, the timing of risks may not be apparent. One aspect of temporal change is captured by a changing failure rate, e.g., the "Bathtub" curve (Figure 2) that specifies a general failure rate profile over system life (Modarres & Groth, 2023).

The bathtub curve is useful for describing the average failure behavior of many identical components or systems overtime – relatively high at the beginning of life as manufacturing defects are worked out of the system, followed by a relatively constant failure rate during the system's "useful life" before the failure rate begins to increase due to aging and wear-out mechanisms. The behavior of a specific system can be modeled using a probability distribution to identify the probability of failure before a specific time (e.g., before the system's design/mission life). Figure 3 displays this sort of "time-to-failure" or "life" model for a system.

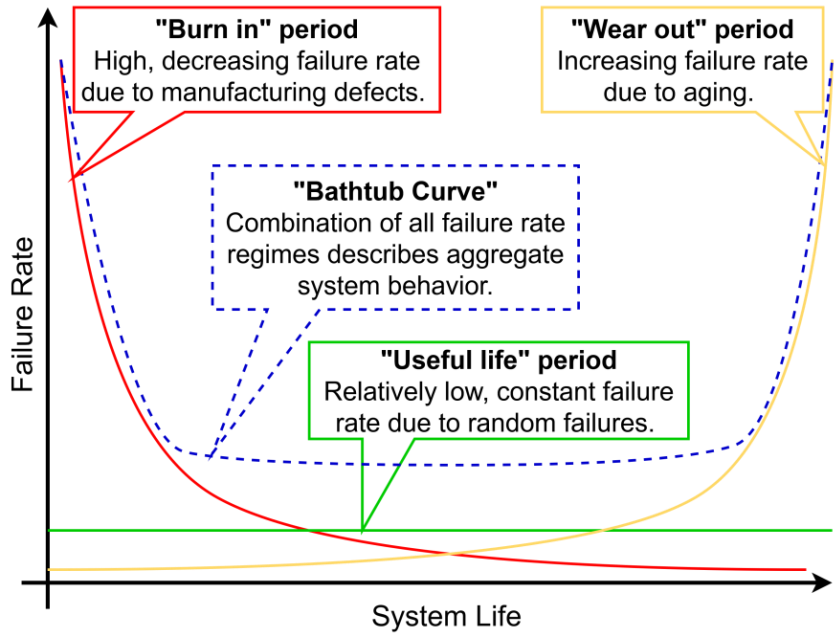

Figure 2. The Bathtub Curve describes how an average failure rate (for system, component, etc.) changes over time due to changing failure behaviors.

As Figure 3 shows, risk is deeply connected to reliability; as the probability of non-failure, reliability is the complement to the probabilistic component of risk. They are not strictly equal, as risk includes a measure of consequences while reliability does not. So, where risk characterizes the failure of a system, reliability addresses the success criteria for a system.

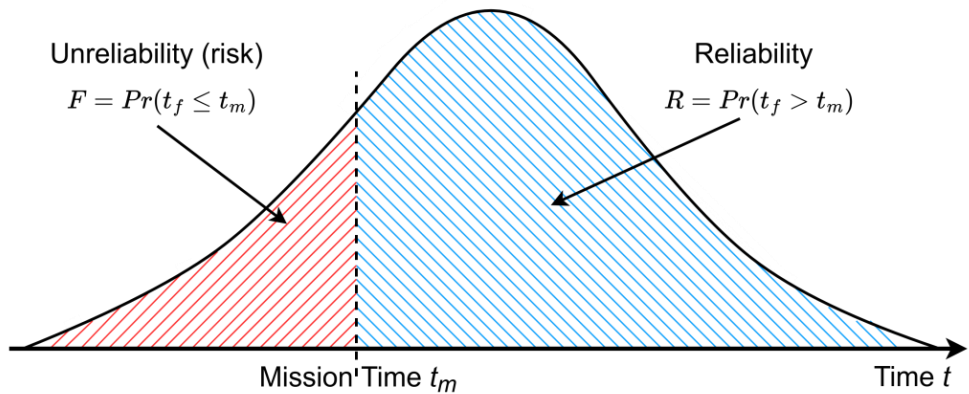

Figure 3. Risk and Reliability as probabilistic metrics in a time-to-failure distribution.

Risk can be described as a function of the system as a whole, of the constituent subsystems, or even of the components. The basis of modern risk analysis – both qualitative approaches like failure modes and effects analysis (FMEA) and quantitative methods like probabilistic risk assessment (PRA) – is that component-level failures propagate to produce system-level effects. Accordingly, many approaches to risk analysis are component-driven, while resiliency assessments may focus more at the subsystem or system level.

## 2.2. Reliability

Reliability, as shown in Figure 3, defines the probability that a system (or subsystem, component, etc.) will perform its designed mission under stated conditions for the desired lifetime (Modarres & Groth, 2023). That is, the reliability of a system evaluated at the mission life, $R_{sys}(t_m)$, is the probability of system *non-failure* over the mission life. The reliability (success-space) statement can be translated into a risk (failure-space) statement by leveraging the proportionality of risk and reliability, shown in Equation (3). The total risk of the system is the probability of failure before the mission life multiplied by the consequence severity.

$$R_{sys}(t) \propto 1 - F_{sys}(t) = 1 - \sum_{i=1}^{n} C(s_i; t) \times \Pr(s_i; t) \qquad (3)$$

Reliability is, like risk, an uncertain system characteristic that is ultimately quantified by a probability distribution. Further, reliability is often explicitly temporal in the sense that the *time* of system failure is critical to determining reliability. Thus, many reliability metrics center around time e.g., the useful life of a system and/or the mean time to (or between) failures (MTTF or MTBF, respectively). The reliability of a system at time $t$ is the probability that a failure will occur *after* that time, as per Equation (4).

$$R(t) = \Pr(t_f > t) = 1 - F(t) \qquad (4)$$

Reliability is, like risk, a well-defined and quantifiable metric of system performance. Many approaches have been developed to determine the reliability of even complex systems. However, as systems grow to the extreme of complexity, with multiple interacting components across domains and tight couplings to other systems, component-based reliability approaches start to fail (Aalund & Paglioni, 2025b). While system-level reliability approaches are being developed to resolve these issues, e.g., (Aalund & Paglioni, 2025a), reliability still struggles to capture dynamic system properties.

More fundamentally, outside of certain, relatively limited, approaches (e.g., repairable system modeling), reliability assessments do not consider failure consequences or system recovery mechanisms. As failures increase in frequency and/or consequences, due to new threat vectors, dependencies, and operating contexts, it is critical that there be a sound, quantitative means of understanding system performance.

## 2.3. Resiliency

Resiliency is defined differently across practitioners, even when limiting discussion to engineering resiliency. Various authors have defined "engineering resiliency" as:

- "The ability to prevent something bad from happening, prevent something bad from becoming worse, or recover from something bad once it has happened" (Boring, 2010).

- "The ability to respond to an unanticipated disturbance […] and then to resume normal operations quickly and with minimum decrement in performance" (Fairbanks et al., 2014);

- "The capability of recovering safely and efficiently from abnormal events" (Patriarca et al., 2018);

- "The capability […] to anticipate, monitor, plan for, respond to, learn from and adapt in the face of unexpected events and maintain system safety" (Paglioni, 2024); *and*

- "Maintaining capability in the face of an adversity" (Brtis et al., 2025).

While clearly different, the selection of resiliency definitions offered above draw on parallel ideas. Firstly, resiliency for engineering systems clearly includes aspects of prevention and response – that is *proactive* resiliency elements and *reactive* resiliency elements. This article will propose a more universal definition for resiliency in section , but broadly resiliency for engineering systems can be identified as the ability to anticipate and respond to perturbations.

The idea of key function and/or structure continuity for systems is central to the broad understanding of resilience as the anticipation of and response to perturbations. In this view, resilience and reliability are related by the basis in a desired and measurable state of system performance. However, reliability and resiliency differ in the approach to defining and measuring functional continuity. Reliability is quantified in terms of the loss of that performance (e.g., a system reliability of 0.9 refers to a probability of performance loss of 0.1), while resiliency is measured based on the *return* to performance, although no unified measurement of this return exists.

Resiliency, as understood for engineering systems, is adapted from ecological resilience, established by Holling in 1973. This seminal work established the core elements of resilience as the evolution of a system around stable domains, and the ability of the system to absorb changes to state and driving variables, and parameters, and still function (Holling, 1973). These elements – of evolution and absorption – laid the foundation for anticipating system transformation in the face of novel threats, and thus paved the way for system prognostics. In this original approach to resiliency, "stability" describes system performance at or near an equilibrium (Pimm, 1984), whereas resilience was developed by Holling to describe system behaviors to move towards an equilibrium.

Figure 4 displays a major split in engineering and ecological resiliency approaches. Classically, engineering resiliency (Figure 4, Left) describes the system actions and time required to return to the steady state of a single equilibrium following some perturbation (Neubert & Caswell, 1997). The characteristics of engineering resilience are described by the "bowl" geometry (e.g., the possible perturbation end states around the single, central equilibrium) and the time required for the system (ball) to return to equilibrium (Scheffer et al., 1993). This was termed "engineering resilience" (Holling, 1996) because this concept – of evolution around a single equilibrium – aligns closely with engineering approaches to designing and maintaining some optimal system structure. Conversely, a second resiliency paradigm, ecological resilience (Figure 4, Right) allows for multiple stable equilibria. In a multi-equilibria system, resiliency can be measured by the magnitude of perturbation(s) that can be absorbed before the system transitions to a new equilibrium. This multi-equilibria approach mirrors ecosystem state transitions found in nature; natural systems are not generally considered to possess only one stable architecture (Holling, 1996).

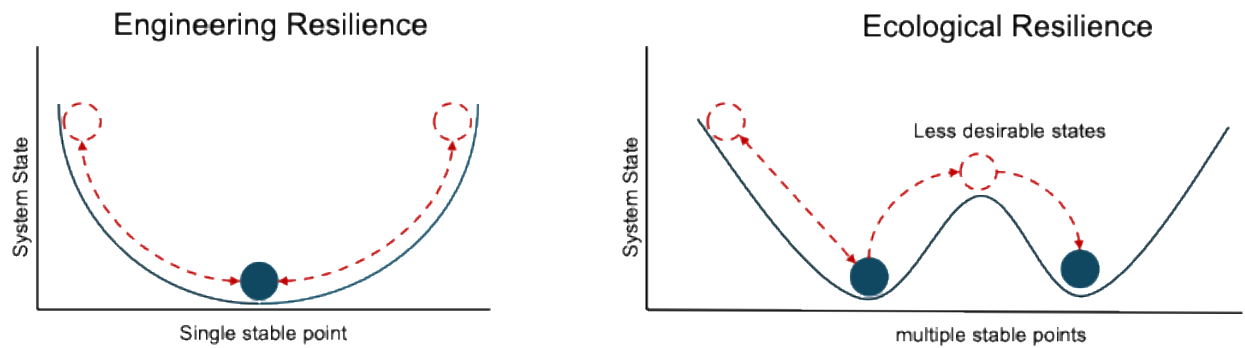

Figure 4. Single vs multiple stable points as views on resilience.

The existence of multiple points of stability (equilibria), and the ability of the system to transition between these islands of stability, is the major differentiating factor between engineering and ecological resiliency approaches. This split affects the definition of stability for the system and how stability, and thus resiliency, can be measured. For example, considering only a single state then necessarily defines "stability" as the forces (internal and external to the system) maintaining the system in proximity to the single equilibrium state. Stability is then measured in terms of the time to return to equilibrium and the system behaviors that affect this return (Bodin & Wiman, 2004). If the system is instead assumed to have multiple stable equilibria, then "stability" is defined by the forces that maintain the system in equilibrium *or* contribute to the transition to a new equilibrium (following a perturbation). Measuring stability in this multi-equilibria approach is more complex. While the time required for the system to achieve stability can still be measured, the system architecture evolution must be considered as part of resiliency. That is, the transition between equilibria may include structural and/or behavioral changes in the system. The process of system change, through periods of growth, conservation, collapse, and reorganization, is the "panarchy" cycle (Gunderson & Holling, 2003).

Resiliency engineering was developed as a means of overcoming the limitations of traditional, static risk and reliability assessments by addressing the dynamic nature of systems (Woods & Wreathall, 2003). In the domain of resiliency engineering, building the capacity to manage unpredictable disturbances is a means to improving system reliability (Hollnagel et al., 2006; Yodo & Wang, 2016). Functionally, this means that resiliency engineering is often focused on *unexpected* perturbations, including those that fall below the "credibility threshold" for typical risk and reliability assessments (e.g., "black swan" events). That is, as Figure 5 shows, reliability and resiliency engineering may focus on separate events.

Despite the growth of resiliency engineering to multiple domains, including many high-consequence, low-probability applications, there are still foundational shortcomings in defining, modeling and quantifying resiliency (Patriarca et al., 2018). Further, although resiliency engineering improves reliability by accounting for dynamic and adapting system pressures, the focus is on improving the recovery time towards a single stable equilibrium, rather than allowing the adaptation of the system and transition to new equilibria.

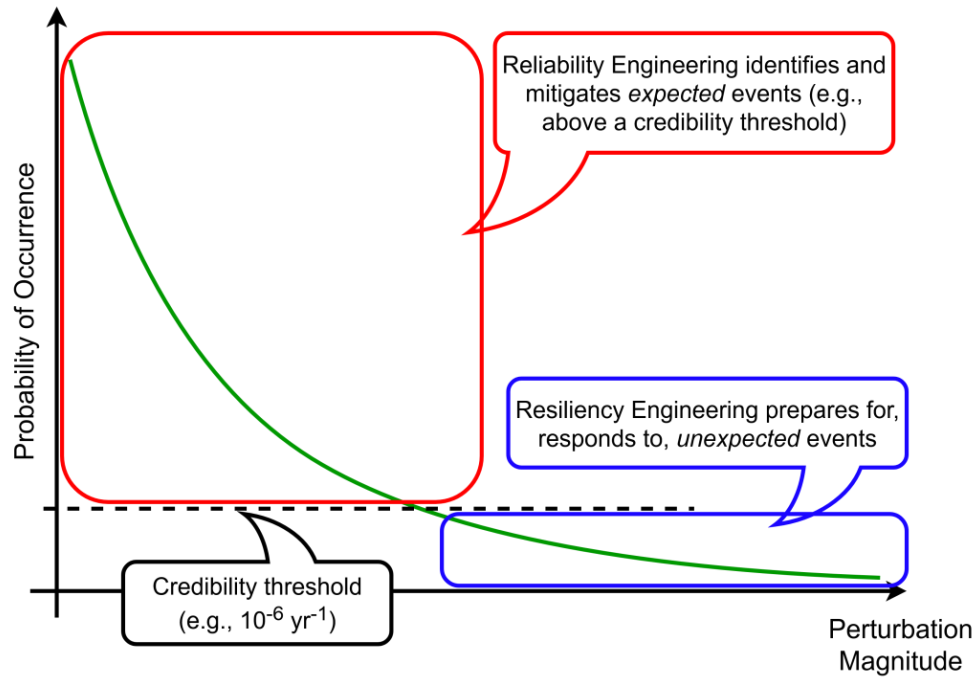

Figure 5. Reliability and resiliency engineering tend to focus on different events. Modified from (Sweetapple et al., 2018).

Resiliency thus improves on typical risk and reliability assessments by adding elements of perturbation recovery. However, resiliency engineering approaches fail to explicitly capture or model the ability of systems to adapt and recover to a new state. Further, resiliency remains a poorly-defined characteristic that is difficult to measure.

## 3. Reliability to Resiliency: Adapting Systems Engineering to Systems

As discussed in the previous section, reliability and resiliency are closely related approaches to understanding system performance and safety. System complexity, number of interconnections, and the consequences of failure are increasing rapidly and outpacing the ability of reliability approaches to adequately describe the system. Further, as systems face novel threats across different vectors, including increasingly frequent (and more severe) natural hazards and cyber hazards, an inability to model and quantify system performance during and after perturbations (threat realizations) is placing critical systems at risk. However, the inclination to shift rapidly from a reliability-centered to a resiliency-centered systems engineering paradigm may not be the best path forward.

### *3.1. Tradeoffs Between Reliability and Resiliency*

As discussed in Section 2, reliability and resiliency are interdependent but separate approaches to system safety. The pursuit of either in isolation can jeopardize the achievement of the other. For example, high-reliability systems may be brittle or fragile in the face of unexpected (i.e., unmodeled) perturbations in operation. The Space Shuttle was designed as a reliable system, was ultimately unable to adapt to changing conditions. The pursuit of reliability in the Shuttle design resulted in more than 3,000 single points of failure items (Marais et al., 2004), all but precluding the resulting design from being resilient. Furthermore, the mental model enforced by the *pursuit* of reliability may have masked the importance of adaptation and recovery mechanisms.

The preceding discussion highlights that the tradeoffs between reliability and resiliency are not limited to the technical domain; indeed, they can bleed into the social organization as well. That is, building a highly-reliable system may jeopardize the system's ability to adapt in the face of unexpected events. The single-minded pursuit of reliability, as the avoidance of failure, may blind designers and operators to the importance of adaptation and recovery mechanisms. In the case of the Space Shuttle, the system was brittle to perturbations (e.g., the "O-Ring Problem"), while the NASA enterprise proved unable to adapt and address the problem.

Similarly, resiliency-focused strategies can act to reduce system reliability. Like the complications injected by the single-minded pursuit of reliability, pursuing resiliency affects the final design through both technical issues and organizational complications. Both technical and organizational complications are borne largely out of increased system complexity that results from pursuing resiliency.

Generally, approaches to system resiliency rely on redundant and/or tightly-coupled components (Marais et al., 2004). Redundant components allow a system to maintain functionality in the case of perturbations, while tight couplings (between subsystems, components, etc.) ensure that the system can respond quickly. However, both of these resiliency approaches ultimately increase system complexity. Redundant components increase the number of components and interactions that can contribute to system failure, while tight couplings increase the complexity of the system logic and the speed at which perturbations can propagate through the system. While both of these approaches improve the system's ability to respond to perturbations, they result in a highly complex system that can be susceptible to even "normal" failures. This recognition is foundational to "Normal Accident Theory," that identifies the inevitability of accidents in systems that are highly complex and tightly-coupled (Perrow, 1999).

At the organizational level, the pursuit of resiliency can complicate the analysis of system reliability (Marais et al., 2004). Reliability models of resilient systems may become intractable due to the large number of components, and may not appropriately track the complicated propagation pathways. Thus, reliability models may end up neglecting components and/or logical pathways, allowing smaller risks to fester. Furthermore, the pursuit of resiliency can enforce a mental model in the organization that focuses on mechanisms to facilitate rapid, graceful responses to perturbations. This mental model, while inherently a positive approach to ensuring system safety, may mask the importance of avoiding those perturbations altogether.

Clearly, a single-minded approach to either reliability or resiliency (as currently defined) leaves both the system and organization vulnerable to perturbations. However, it is important to realize that the tradeoffs discussed herein are *extrinsic* conflicts, rather than intrinsic incompatibilities between reliability and resiliency. That is, completely disregarding reliability in favor of resiliency is not an optimal solution. There are synergies that exist between reliability and resiliency that can promote a unified, effective approach to system safety.

## 3.2. *Synergies Connecting Reliability and Resiliency*

The tradeoffs enforced by the pursuit of reliability or resiliency are just that: enforced, rather than inherent. There are theoretical and practical synergies to be obtained by considering reliability *and* resiliency together in system designs. Practically, reliable systems enforce resiliency by avoiding the precipitating failure altogether. At the same time, resilient systems can enforce system reliability *in the long term*, i.e. by allowing the mission to continue even if a subsystem or component has failed. Theoretically, there are quantitative approaches to reliability engineering that can be extended to produce quantitative resiliency methods.

The dual pursuit of reliability and resiliency requires going beyond "simple" redundancy, which is an inefficient method of gaining resiliency and can hamper reliability (Marais et al., 2004). There are, however, "components" in complex systems that can provide the flexibility and adaptability needed for resiliency without sacrificing system reliability: the human operators. Humans are often the "last straw" in a system-level event, and thus tend to carry the burden of system failures. This is perhaps best viewed in the lens of the Swiss Cheese Model (Reason et al., 2006; Larouzee & Le Coze, 2020) and the recognition that humans are blamed for 60-80% of industrial accidents and incidents. However, humans are also the system element best equipped to help the system rebound in the face of unexpected perturbations (Boring, 2009; Paglioni, 2024).

Humans thus provide a practical synergy between reliability and resiliency: designing for human reliability can directly improve resiliency by facilitating adaptability and flexibility. Designing for resiliency can produce lower stress environments that facilitate higher operator reliability. This human-centric approach has a theoretical synergy as well, considering the overlaps between resiliency engineering and human reliability analysis (Boring, 2009, 2010).

Applied examples of the synergies between reliability and resiliency can be found in High Reliability Organizations (HROs). These organizations operate in high-risk (often high-consequence, low-probability) domains but maintain a high level of safety through adherence to several practices and principles (Weick & Sutcliffe, 2007), as shown in Table 1. Broadly identified, HROs are preoccupied with the potential for failure (given the high consequences), reluctant to oversimplify the problem or solution space, maintain deep interconnections across the organizational structure, and actively build resiliency capabilities.

Table 1. Principles and Practices of High Reliability Organizations.

| Principle | Practice | Pathway |
|---|---|---|
| *Preoccupation (with failure)* | Continuous search for *potential* failures and near misses. | Proactive |
| *Reluctance (to simplify)* | Avoid oversimplification; embrace domain complexity. | Proactive |
| *Sensitivity (to operations)* | Connect all organizational levels to operations; Maintain real-time strategic & tactical awareness. | Reactive |
| *Commitment (to resiliency)* | Actively build capabilities for responding to, adapting and recovering from unexpected events. | Proactive |
| *Expertise* | Decision-making charged to relevant expertise, not rank. | Reactive |

The principles and practices employed by HROs in their pursuit of reliability can be delineated into one of two major pathways for reliability. The first is the **proactive** pathway, where organizations commit to identifying and mitigating failures and perturbations before they occur. The second is the **reactive** pathway, where HROs are capable of swift action in the face of unexpected events (Sutcliffe, 2011). That is, reliability-seeking socio-technical systems view resiliency as integral to reliability because of its complementary focus on failure recovery.

The principles and practices of HROs aligned with the *reactive* pathway reinforce some of the preconceived notions of resiliency-building characteristics in systems, outlined in Sections 2.3 and 3.1. The principle of sensitivity to operations is often implemented as a responsive organizational structure, where information is able to freely flow between hierarchy levels and across organization domains. Often referred to as "safety culture," this is the organizational equivalent of a tightly-coupled system: able to respond quickly in the face of perturbations, but with the risk that faults (misinformation, rot) can quickly propagate.

This section outlined the various synergies that exist connecting reliability to resiliency, and further demonstrated that a shift to resiliency-informed decision-making does *not* mean foregoing the assessment and improvement of system reliability. On the contrary, as Figure 1 shows, reliability is a *necessary*, but not *sufficient* aspect of system resiliency. The goal with resiliency-informed decision-making should be to leverage "traditional" risk and reliability approaches, which are well-defined and widely used to identify the cause, probability, and effect of system failures, the remaining useful life, and the anticipated system safety. Augmenting these approaches with resiliency theory and engineering ensures that decision-makers have a robust understanding of all aspects of the problem space, from anticipation (of perturbations) to response and adaptation. The next section discusses one preliminary outcome of the effort to bridge reliability and resiliency.

## 4. Bridging Reliability and Resiliency

As previously discussed, there are significant overlaps between reliability and resiliency. One such overlap is borne from viewing repair as a subset of resiliency-enforcing actions. In this perspective, repairing a system is one approach to reactive resiliency (**Error! Reference source not found.**, 7). This section reviews the foundations of repairable system modeling and extends these methods to capture system resiliency.

### *4.1. Repairable System Modeling*

Repairable system modeling aims to identify system availability as a function of time, considering the potential for failure and repair actions. There are various approaches to modeling repairable systems, each with unique assumptions – largely around the failure distribution and effectiveness of repair. Regardless of method chosen, the variables of interest are typically availability, rate of occurrence of failure (ROCOF), and/or mean time between failures (MTBF). Table 2 briefly reviews available approaches to repairable system modeling. Readers are directed to (Modarres & Groth, 2023, Chapter 7) for a more detailed introduction to repairable system modeling.

Table 2. Overview of stochastic point processes for repairable system modeling.

| **Stochastic Point Process** | **Assumptions and Characteristics** | **Notes** |
|---|---|---|
| **Homogeneous Poisson (HPP)** | Exponentially-distributed times to failure (TTF) with constant ROCOF $\lambda$. Perfect repair assumption. | Memoryless – preventive repairs do not work in this approach. |
| **Renewal Process (RP)** | Generalized HPP with any TTF distribution. Perfect repair assumption. | Closed-form solution only for underlying gamma distribution. |
| **Nonhomogeneous Poisson (NHPP)** | Time-dependent ROCOF $\lambda$. Minimal repair assumption. | Multiple versions with different ROCOF functions (Crow-AMSAA, Cox-Lewis, Linear). |
| **General Renewal Process (GRP)** | Variable repair assumption (parameter of rejuvenation $q$). Any underlying TTF distribution allowed. | |

The repair assumption made in the course of modeling the repair process is critical to understanding the system behavior with time. The *perfect repair assumption* dictates that the repair action resets the component (or system) to an "as-new" state, that is the component's age is reset to zero. Both the HPP and RP assume perfect repairs. The *minimal repair assumption*, on the other hand, means that the component is repaired to the "same-as-old" condition, that is its condition following repair is identical to its condition immediately prior to failure. The NHPP, regardless of underlying distribution employed, follows the minimal repair assumption. Clearly, neither the perfect repair nor minimal repair assumptions are realistic for most systems. The GRP (Kijima & Sumita, 1986) was developed to relax the repair assumptions required in the HPP, RP, and NHPP approaches, and accordingly can cover all those cases, as well as intermediate repairs and "better than new" (i.e., system improvement) conditions.

The GRP builds on the RP by introducing the parameter of rejuvenation, $q$, that governs the effectiveness of the repair. The value of $q$ is not bounded; Figure 6 displays the corresponding repair assumption for different values of the $q$ parameter and a depiction of system state with time under each repair assumption.

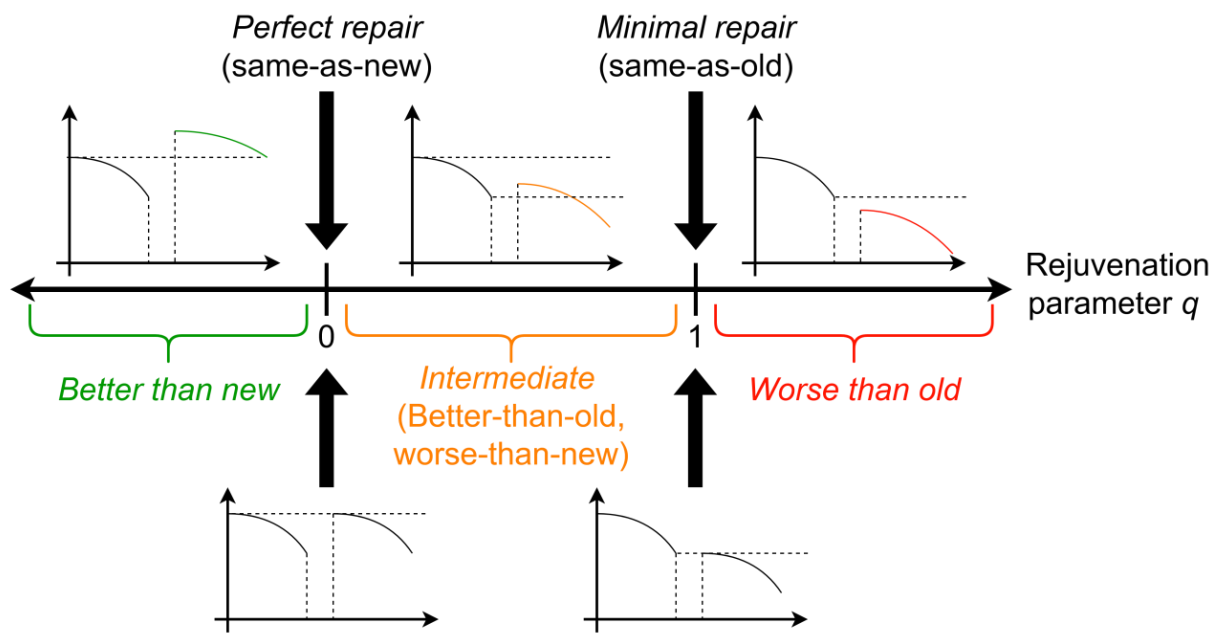

Figure 6. Values of rejuvenation parameter $q$ affect repair assumption and system state over time.

## 4.2. Resiliency Modeling

Logically, repairing a system following a failure is one pathway to (reactive) resiliency. This perspective facilitates the application of repairable system modeling techniques to reactive resiliency. Being able to quantify resiliency for a general system – in any capacity – is an improvement on current practice, where quantification approaches, if available, may be limited to specific applications (e.g., (Panteli et al., 2017)).

In this approach, the parameter of rejuvenation $q$, that dictates the effectiveness of repair, (Kaminskiy & Krivtsov, 2015), is expanded to characterize the total restoration capacity of the resilient system. That is, this quantifies *reactive* resiliency enforced by any action (repair, self-healing, adaptation, etc.). As Table 3 shows, the "resiliency degree" $q_{res}$ follows a similar pattern to the rejuvenation parameter $q$ (Figure 6), but structured to support the goal of quantifying *resiliency* via Equation (5). Table 3 further demonstrates that resiliency can be cast in terms of reliability and risk. Figure 7 demonstrates the effect of resiliency degree on system operation.

Equation (5) provides a first approach to quantifying reactive resiliency ($\rho_r$), considering the time required to effect the resiliency actions ($t_{res}$) and degree of resiliency $q_{res}$. In this approach, $q_{res}$ defines the system reliability gained via resiliency actions; That is, $q_{res}$ is the difference between as-new reliability (assumed $R = 1$) and the system reliability following resiliency. The times $t_{res}$, $t_{fail}$, and $t_{mission}$ are the time required to effect resiliency actions, time of failure, and mission lifetime, respectively. Thus, the quantified reactive resiliency is responsive to two main characteristics of resiliency: *effectiveness* of actions and the *time* required to perform them.

Table 3. Resiliency degrees

| Resiliency Degree | Reliability Definition | Risk Definition | Figure 7 |
|---|---|---|---|
| Better than new $q_{res} < 0$ | System is repaired to a state better than when system was new. | System risk lower than when system was new. | Recover to Line (1) |
| Good-as-new $q_{res} = 0$ | System is resilient such that reliability following repair R=1. | System risk equal to when system was new. | Recover to Line (2) |
| Same-as-old $q_{res} > 0$ | System reliability following repair equivalent to its reliability immediately prior to failure. | System risk is equal to immediately prior to failure. | Recover to Line (3) |
| Worse-than-old $q_{res} > 0$ | System reliability following repair lower than immediately prior to failure. | System risk is higher than immediately prior to failure. | Recover to Line (4) |

$$\rho_r = (1 - q_{res}) \cdot \frac{t_{mission} - t_{fail} - t_{res}}{t_{mission} - t_{fail}} = (1 - q_{res}) \cdot \left(1 - \frac{t_{res}}{t_{mission} - t_{fail}}\right) \quad (5)$$

The time required to perform resiliency actions, $t_{res}$, may take any value (i.e., $t_{res} \in [0, \infty)$). As $t_{res} \to 0$, resiliency approaches the resiliency degree: $\rho_r \to (1 - q_{res})$. Thus, a system that instantaneously recovers (i.e., $t_{res} = 0$) to a good-as-new state ($q_{res} = 0$) is "perfectly resilient," and $\rho_r = 1$. If the time required to effect resiliency actions exceeds the mission time, i.e., the system is not recovered prior to the anticipated end of life, then system resiliency $\rho_r = 0$, and the system is "perfectly non-resilient."

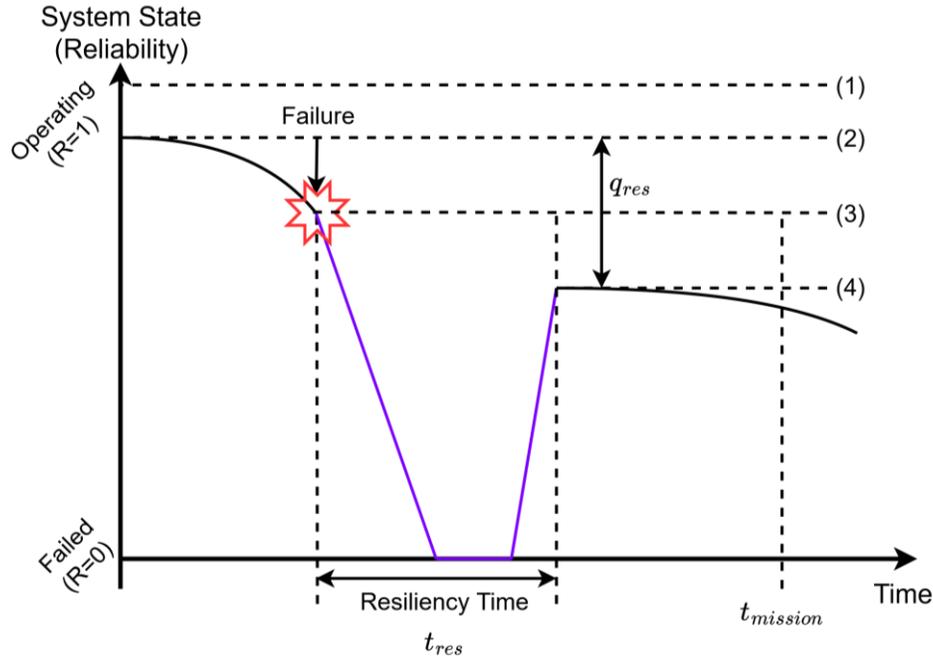

Figure 7. System resiliency visualization.

This quantification of reactive resiliency in terms of recovery time and recovery degree reframes the conceptual relationship between reliability and resiliency. Reliability is, in this view, a first-order property: the probability that a system will not fail under specified conditions. Resiliency, by contrast, is the reliability of a system's capacity to recover from unreliability. Put differently, resiliency is understood here as a second-order property, or meta-condition, that becomes salient only once reliability has been lost. Once a system has entered into disruption, resiliency functions as the intra-disruption measure of system health. The level of operation regained following perturbation can be expressed as a recovered reliability. In this sense, resiliency spans both conceptual registers: it is meta-reliability in the abstract, and it is the operative measure of performance in the midst of, and following, disruption.

This dual character also reveals a temporal asymmetry between the two properties. Reliability predominantly governs the pre-failure horizon, where the relevant question is whether the system will avoid disruption within its mission life. Resiliency predominantly governs the post-failure horizon, where the relevant question is how the system behaves once disruption has occurred. Within its post-failure horizon, resiliency provides a continuous account of system health across the disruption interval, capturing both the trajectory of degradation and the degree of recovery. In this respect, reliability is a prospective, probabilistic measure anchored in the avoidance of failure, whereas resiliency is a dynamic, processual measure anchored in the system's recovery trajectory.

These temporal differences between reliability and resilience correspond to distinct ontologies of system health. Reliability presumes a binary state space: a system either succeeds in avoiding failure, or it does not. Resiliency, by contrast, is better captured as a graded continuum of system states across disruption, ranging from degraded operation through partial recovery to full or even improved function. The introduction of resiliency degrees in Table 3 makes this continuum explicit: recovery can leave the system worse-than-old, same-as-old, as-good-as-new, or better-than-new. By situating recovered reliability within this spectrum, resiliency extends the binary logic of reliability into a more nuanced account of performance that captures the quantitative differences among recovery trajectories, rather than collapsing them into a simple dichotomy of success and failure.

Taken together, these distinctions carry a normative implication for system design and decision-making. In relatively simple or isolated systems, reliability may suffice as the primary basis for evaluation, since the principal concern is to minimize the likelihood of failure. In complex, tightly coupled, and interdependent systems like infrastructure, however, failures are not exceptional anomalies but inevitable features of system behavior. Here, resiliency provides the more appropriate basis for judgment by evaluating not only whether failures can be avoided, but *also* how systems absorb, adapt to, and recover from them. Reliability secures stability under expected conditions; resiliency secures continuity once conditions depart from expectation. For systems where disruption is unavoidable, resiliency becomes a better measure of system health.

## 5. Discussion

The persistent challenge in advancing resiliency from an appealing concept to a practical tool lies in its lack of a clear, measurable basis. Across infrastructure sectors, resiliency is invoked as a critical property of system health, yet its meaning often remains diffuse: sometimes equated with reliability, sometimes reduced to redundancy, and sometimes treated as a matter of organizational culture or coordination. Without consistent definitions and quantifiable measures, resiliency risks remaining aspirational, difficult to compare across systems, and difficult to embed into decision-making frameworks that must weigh cost, risk, and other tradeoffs under real-world constraints.

The importance of moving beyond this ambiguity was reinforced by recent Colorado whole-of-state infrastructure discussions that brought together representatives from water, wastewater, dams, energy, communications, and transportation sectors, along with state and federal partners (Hunyadi et al., 2025). Despite their diverse mandates, participants converged on several key themes: the fragility of critical interdependencies, the central role of cybersecurity in modern infrastructure, and the practical challenges of coordination across jurisdictions and asset classes. Resiliency emerged as a unifying concern and point of interest, but the very prominence of the term underscored its current limitations. Attendees repeatedly highlighted the difficulty of translating the idea of resiliency into operational terms that could guide investment priorities, emergency response planning, and long-term system design.

This need is not confined to one state or conference. Modern infrastructure is characterized by tight coupling, complex interdependencies, and exposure to novel threats. Failures propagate across sectors with increasing speed, producing system-of-systems behaviors that defy siloed planning or design. In such contexts, the value of resiliency is precisely its attention to post-failure dynamics: not whether disruptions will occur, but how systems can absorb, adapt, and recover once they do. Yet for resiliency to inform decisions in this landscape, it must be rendered actionable. That means defining metrics that can be consistently applied across sectors, compared across competing designs, and integrated into regulatory and management frameworks alongside established measures of risk and reliability.

The resiliency framework advanced in this paper should be understood as a step toward that goal. By quantifying reactive resiliency in terms of recovery time and recovery degree, it provides an initial metric that connects naturally to reliability engineering while addressing the behaviors that matter most during and after disruption. This is necessarily a partial account; it captures only reactive resiliency, leaving proactive dimensions such as anticipation, planning, and adaptation for future work. Nevertheless, even this partial account demonstrates the feasibility and value of rendering resiliency measurable. It offers a concrete starting point for integrating resiliency into the models and assessments that guide infrastructure planning and emergency preparedness.

# 6. Conclusions

Reliability and resiliency are clearly intertwined system characteristics that can feed into and/or off of the other. For example, the quantification approach laid out in Section 4.2 casts resiliency in terms of reliability, demonstrating that there can be positive interactions between them. On the other hand, optimizing a system for either reliability or resiliency may degrade the other, as discussed in Section 3.2. Viewing reliability and resiliency as interconnected allows multi-objective approaches to system safety and avoids sacrificing reliability in favor of resiliency, or vice-versa.

This paper provided an overview of the current understanding of reliability and resiliency, and demonstrated an available approach to quantifying resiliency by extending existing reliability techniques. However, operationalizing resiliency for system-level decision-making, and particularly for critical infrastructure, requires a more detailed understanding of resiliency. Principally, as evidenced by Section 2.3 and the discussions at the recent Interlock 2025 summit, there is a clear need for a common terminological basis around resiliency. **Establishing a coherent definition of resiliency in the context of engineered systems is a prerequisite** for further study of resiliency and implementation into any decision-making framework.

Additionally, the mathematical approach outlined here in Section 4.2 needs to be validated and further expanded to consider *proactive* resiliency. **A robust quantitative approach to resiliency is integral** to developing and refining modeling architectures that can support both reliability and resiliency. Finally, operationalizing resiliency for critical systems means developing a supportive decision-making framework that can compare resiliency measures, inter-system resiliency and reliability characteristics, and predict the impact of proactive resiliency actions. This requires **implementing reliability and resiliency approaches into a common, robust modeling architecture** (e.g., Bayesian Networks, Markov Models, etc.) and developing the "resiliency-informed decision-making" approaches suitable for critical systems.

Future work in this area will focus on the areas of need identified above, beginning with developing a common terminological basis for resiliency in the context of engineering systems. This means defining resiliency as a construct and characteristic, as well as the facets of resiliency and categories of resiliency actions. While future work is done in the context of critical systems (i.e., lifeline infrastructure), the goal is to produce a resiliency-focused engineering paradigm that is suitable for all engineered systems, and harmonized with resiliency approaches from other fields (e.g., ecological resiliency).